# Impact of spontaneous emission on spin-squeezed quantum sensors

Jinyang Li[1] and Selim M Shahriar[1,2]

[1]*Department of Physics and Astronomy, Northwestern University, Evanston, IL 60208, USA*

[2]*Department of Electrical and Computer Engineering, Northwestern University, Evanston, IL 60208, USA*

## Abstract

The echo squeezing protocols (ESPs) are techniques that amplify the phase shift in a quantum sensor with one-axis-twist squeezing (OATS). For atomic sensors, spontaneous emission (SE) in the OATS operations is an important imperfection, which, however, is prohibitively difficult to study. SE can transfer atoms to all the Zeeman substates, making the number of relevant collective states excessively large. In this paper, we focus on a relatively simple isotope, namely $^{88}$Sr, which only has one Zeeman substate in each of the two ground states. Nevertheless, studying the effect of SE is still challenging because SE will populate all collective states, either symmetric or asymmetric, thus putting the ensemble into a mixed state. Based on analytical derivation and numerical simulations, we conclude that the GESP is more resistant to SE than the conventional echo squeezing protocol (CESP). This is another advantage of the GESP that has not been realized formerly. We also find that for the GESP employing $^{88}$Sr, the SE-induced reduction in the signal contrast is the same as that for a Ramsey protocol without squeezing and the quantum noise is suppressed by SE. This is an unexpectedly favorable result. The SE-induced suppression of quantum noise constitutes a previously unrecognized effect that challenges both earlier conclusions and intuitive expectations.

## 1. Introduction

Spin squeezing [1] is a promising technique for pushing quantum sensors towards the Heisenberg limit. One-axis-twist squeezing (OATS) is the most experimentally accessible type of spin squeezing. For concreteness, this paper is restricted to only OATS. The OATS is described by the Hamiltonian $\hbar\chi S_z^2$ and the corresponding propagator $e^{-i\mu S_z^2}$, where $S_z$ is the $z$ component of the dimensionless spin operator representing the sum of the spin operators for all the atoms, and $\chi$ is represents the strength of the non-linear interaction needed for producing the squeezing effect. The effect of the OATS is characterized by the squeezing parameter, $\mu \equiv \chi T$, where $T$ is the duration of the squeezing interaction.

The resulting state after an OATS operation is applied to a coherent spin state oriented in the $x$ direction is as follows. As $\mu$ increases from zero, the uncertainty of the spin operator in a certain direction decreases at the expense of an increase in the uncertainty along a perpendicular direction. As $\mu$ continues to increase, the resulting state becomes more complicated. For special cases when $\mu = \pi/n$, the resulting state is a superposition of $n$ coherent spin states evenly distributed around the equator of the Bloch sphere [2]. More specially, for $\mu = \pi/2$, the resulting state is a Schrödinger cat state consisting of two coherent spin states oriented in opposite directions. The OATS can be used for magnifying quantum phase shift (i.e., narrowing the signal fringes as a function of the phase shift) if it is applied in the echo configuration. The protocols in such a configuration are named echo squeezing protocols (ESPs). These ESPs in principle can be adapted to any atomic sensor employing coherent quantum states, for example, Ramsey clocks [3], Ramsey magnetometers [4], and atom interferometers [5, 6].

There are three types of ESPs, which produce different degrees of phase magnifications and noise amplifications. The first one is the conventional ESP (CESP) [7, 8], which can ideally magnify the phase shift by the factor of $\sqrt{N/e}$ ($N$ being the number of atoms), while leaving the quantum noise unchanged, for a particular value of $\mu$. The second one is the generalized ESP (GESP) [9], which can ideally magnify the phase shift by the factor of $N \sin \mu / \sqrt{2}$, while amplifying the quantum noise level by the factor of $\sqrt{N} \sin \mu$, for a broad range of values of $\mu$. The final one is the Schrödinger cat state protocol (SCSP) [9, 10], which can ideally magnify the phase shift by the factor of $N$, while amplifying the quantum noise level by the factor of $\sqrt{N}$, for $\mu = \pi / 2$, if the parity (i.e., whether $N$ is even or odd) is known a priori.

In what follows, the spin-1/2 spinors specifically refer to two-level atoms. There are primarily two methods for implementing spin squeezing of atoms, namely cavity-mediated [11, 12, 13, 14, 15] and Rydberg-mediated [16, 17, 18] nonlinear interaction. For either method, spontaneous emission (SE) is an inevitable imperfection. For concreteness, we focus on cavity-mediated nonlinear interaction in this paper. Existing analyses regarding the effect of the SE on squeezing only focus on the increase in the noise level induced by SE, because the original purpose of squeezing was to suppress the quantum noise. The impact of SE on the degree of coherence has been widely ignored. However, to make a squeezed state useful for sensors, taking into account the effect of SE on the degree of coherence is also essential, because it determines the signal contrast at the end of an experimental shot. Moreover, in many cases, SE does not increase the noise level, and even decreases the noise level in some scenarios, which will be discussed in detail later. In such cases, the only negative effect of SE is to decrease the signal contrast. Accordingly, decreasing the signal contrast is in some cases a more important effect of SE than increasing the noise level. There are two simple examples when SE does not increase the noise level. The first example is an atomic clock without squeezing. Such a clock is normally operated at the point of the maximum signal gradient with respect to the frequency. At this point, each atom is in an equal superposition of the spin-up and spin-down states, and the variance of either the spin-up or the spin-down state population is maximized if there is no entanglement. In this case, no factor, including SE, can increase the variance further, if the total number of the atoms is fixed. A similar example is the SCSP. For such a protocol, the variance of the measured operator is maximized among possible states with the same number of atoms, either entangled or untangled. Therefore, no factor, including SE, can increase the variance further.

It is extremely difficult to study the impact of SE on the signal contrast for alkali atoms. It should first be noted that the hyperfine ground states of an alkali atom consist of multiple Zeeman substates, with the $m_F = 0$ states corresponding to the two states in the two-level model. SE can transfer an atom that is entangled to other atoms to an $m_F \neq 0$ state. The atoms transferred to the $m_F \neq 0$ states are still subject to the cavity field, and the whole system will become highly mixed. Therefore, in this paper we study the effect of SE with $^{88}$Sr atoms because $^{88}$Sr only has the $m_F = 0$ Zeeman sublevels in the states $^1S_0$ and $^3P_0$. To implement OATS in this system, the cavity field couples the $^1S_0$ state to the $^3P_1$ state, and the $^3P_0$ is not coupled to other states. The cavity field is far detuned from the $^1S_0 \leftrightarrow {}^3P_1$ transition, with the detuning denoted by $\Delta$. After adiabatic elimination of the $^3P_1$ state [19], the quantum state of each atom is reduced to a two-level system. This reduced system has only one dissipation channel, which is the dephasing of the $^1S_0$ state at a rate $\Gamma = \Gamma_0 \Omega^2 / 4\Delta^2$, where $\Gamma_0$ is the SE rate of the $^3P_1$ state and $\Omega$ is the Rabi frequency induced by the cavity field. It is shown in Appendix B of Ref. [4] that $\chi = \frac{\Omega^2}{4\Delta} \frac{(2g)^2}{4\Delta} \frac{\delta}{\left(\delta^2 + \kappa^2/4\right)}$, where $2g$ is the Rabi frequency induced by a single photon in the cavity field, $\delta$ is the detuning of the probe beam from the cavity resonant frequency, and $\kappa$ is the linewidth of the cavity.

Even if the atoms assumed to be used are clean two-level systems, the study is still challenging. In the presence of SE, the quantum states with all possible total spins can be populated, in contrast to the ideal case where the atoms are only restricted to the $S = N/2$ manifold. Accordingly, the dimension of the relevant Hilbert space is increased from $N+1$ to $2^N$. Moreover, because SE puts the atoms in a mixed state, a density operator instead of a state vector is necessary to describe the quantum state. Fortunately, when the dissipation channels only include dephasing, which does not change the population distribution, the SE-induced reduction in the signal contrast can be calculated analytically for the GESP. We find that the signal contrast decreases to $e^{-\Gamma t/2}$ (the ideal signal contrast being unity), regardless of the value of $N$. Here, $t$ is the total duration of the OATS operation and its inverse. Simulations with the values of $N$ permitted by available computational resources show that the signal contrast decreases to $e^{-3\Gamma t/4}$, regardless of the value of $N$.

The existing investigation into the impact of SE on the quantum noise is very limited. Ref. [12] provides an estimate of the increase in the variance of the spin operator along the short axis of the noise ellipse in the regime of moderate squeezing where the variance along the short axis of the noise ellipse still decreases with increasing $\mu$. This estimation completely ignores the atoms transferred to $m_F \neq 0$, even though alkali atoms are assumed to be used. Ref. [7] directly cites this result to estimate the increase in the noise level for the CESP. Even if the accuracy of the cited result is not considered, the validity of applying this result to the CESP is still questionable because in the CESP, OATS is operated in the oversqueezed regime (in contrast to the moderate squeezing regime). Ref. [10] estimates the increase in the noise level with the random walk model for the SCSP. This result is directly inconsistent with the discussion above that the noise level is maximized and cannot be increased further by SE. We show analytically that when the dissipation channels only include dephasing, the quantum noise level is not affected by SE. While the zero increase in the noise level is already encouraging, our investigation with numerical simulation reveals an unexpectedly favorable result: SE decreases the noise level for the GESP and the SCSP. Given that SE also decreases the signal contrast less for the GESP, it can be concluded that the GESP is more resistant to the CESP. This is another advantage of the GESP compared to the CESP that has not been found previously.

## 2. Model describing SE in the OATS process

In the process of squeezing $^{88}$Sr atoms, the only dissipation channel is the dephasing of the $^1S_0$ state. The SE of the $j$ th atom ( $j=1,2,\cdots,N$ ) is described by the Lindblad operator $L_j=\sqrt{\Gamma}|1\rangle_{j\,j}\langle 1|$, where $|1\rangle$ represents the $^1S_0$ state and $|0\rangle$ the $^3P_0$ state. They correspond to the spin-up and spin-down state in the spin model such that $s_{jz}\left|{}^1_0\right\rangle=\pm\frac{1}{2}\left|{}^1_0\right\rangle$. The master equation for the OATS process can be expressed as

$$\dot{\rho}=-i\left[\chi S_z^2,\rho\right]+\sum_{j=1}^{N}\left(L_j\rho L_j^\dagger-\frac{1}{2}\left\{L_j^\dagger L_j,\rho\right\}\right) \tag{1}$$

where the term $-\frac{1}{2}\left\{L_j^\dagger L_j,\rho\right\}$ describes the decrease in the absolute values of the density-matrix elements (and thus referred to as the dissipation and decay term) and the term $L_j\rho L_j^\dagger$ describes the increase in the absolute values of the density-matrix elements (and thus referred to as the source term). Eq. (1) is a basis-independent differential equation governing the system. To implement explicit calculation, it is necessary to choose a basis. Given that the system is not restricted to the $S=N/2$ manifold, the most convenient basis is the product basis. To illustrate the algorithm, we use the $N=3$ case as an example. The product basis includes eight states, namely $|000\rangle$, $|001\rangle$, …, $|111\rangle$. Obviously, the digits in the ket notation can be considered as binary numbers. For convenience, a product state is denoted by $|a\rangle$, where $a$ is the decimal value of the corresponding binary number. For the state $|a\rangle$, the number of atoms in the $|1\rangle$ state is denoted by $N_1(a)$. For example, we have $N_1(5)=2$. Each term on the right side of Eq. (1) is expressed in the product basis as follows.

$$\langle a|-i\left[\chi S_z^2,\rho\right]|b\rangle=-i\left\{\left[N_1(a)-\frac{N}{2}\right]^2-\left[N_1(b)-\frac{N}{2}\right]^2\right\}\langle a|\rho|b\rangle \tag{2}$$

$$\langle a|\frac{1}{2}\left\{\sum_{j=1}^{N}L_j^\dagger L_j,\rho\right\}|b\rangle=\frac{\Gamma_0}{2}\left[N_1(a)+N_1(b)\right]\langle a|\rho|b\rangle \tag{3}$$

$$\langle a|\sum_{j=1}^{N} L_j \rho L_j^\dagger |b\rangle = \Gamma N_1(a\wedge b)\langle a|\rho|b\rangle \tag{4}$$

where $a\wedge b$ is the result of bitwise AND in the decimal base. For example, $3\wedge 6=(2+1)\wedge(4+2)=4(0\wedge 1)+2(1\wedge 1)+(1\wedge 0)=2$. It should be noted that in the single-atom model used to describe unentangled atoms, the source term shown in Eq. (4) is nonzero only if $a=b$, meaning that the source term does not contribute to the coherence elements in the density matrix. However, the source term in this model does contribute to coherence elements (the elements with $a\neq b$ ) because $|a(b)\rangle$ is a product state while $L_j$ only acts on one atom. Substituting Eqs. (2), (3), and (4) into Eq. (1) yields

$$\begin{aligned}\langle a|\dot{\rho}|b\rangle &= \left\{ i\left[N_1(b)-\frac{N}{2}\right]^2 - i\left[N_1(a)-\frac{N}{2}\right]^2 - \frac{\Gamma}{2}\left[N_1(a)+N_1(b)\right]+\Gamma N_1(a\wedge b)\right\}\langle a|\rho|b\rangle \\ &\equiv (Q_{ab}+D_{ab})\langle a|\rho|b\rangle\end{aligned} \tag{5}$$

where

$$Q_{ab} \equiv i\left[N_1(b)-\frac{N}{2}\right]^2 - i\left[N_1(a)-\frac{N}{2}\right]^2 \tag{6}$$

is the coefficient describing the effect of squeezing, and

$$D_{ab} = -\frac{\Gamma}{2}\left[N_1(a)+N_1(b)\right]+\Gamma N_1(a\wedge b) \tag{7}$$

is the coefficient describing the effect of dissipation. The solution to Eq. (5) is given by $\langle a|\rho|b\rangle = e^{(Q_{ab}+D_{ab})t}\langle a|\rho|b\rangle = e^{Q_{ab}t}e^{D_{ab}t}\langle a|\rho|b\rangle$. This solution implies that although squeezing and SE occur simultaneously, the effect is equivalent to atoms undergoing them sequentially.

## 3. Effect of SE on the signal contrast and the quantum noise level for the ESPs

The simulation results for the effect on the signal contrast and the quantum noise level are shown in this section. Before discussing the result for each ESP, the protocol is briefly summarized. The steps of the CESP are described below. Atoms are initially prepared in the CSS oriented in the $x$ direction. Step 1: An OATS operation is applied. Step 2: A phase shift $\phi$ is introduced, represented by a rotation around the $y$ axis. Step 3: The inverse of the previous OATS operation is applied. Step 4: The spin operator $S_y$ is measured. Figure 1 shows the signal and the quantum noise level for $N = 5$ and $\mu = \mathrm{arccot}\sqrt{N-2} = \pi/6$ (the optimal value of $\mu$ for the CESP). The blue solid curves show the results for $\Gamma = 0.2\chi$ and the red dashed for $\Gamma = 0$. The vertical black dashed line marks the point of the maximum phase gradient, which is the phase shift at which the protocol should be operated. The signal contrast is reduced and the noise level at the operation point is not affected by SE.

The validity of this encouraging result for arbitrary $N$ can be proven analytically. Recalling SE can be treated as occurring after the OATS operation, the process for $\phi = 0$ is equivalent to SE following the OATS operation and its inverse. Obviously, the inverse operation cancels the effect of the OATS process. Therefore, the whole protocol is equivalent to a Ramsey protocol without squeezing. Because the only effective dissipation channel is the dephasing that does not change the population distribution, the noise level will not be affected by SE. The GESP is the same as the CESP except the operator measured in the end is $S_x$. This small difference in the steps results in a huge difference in behavior. For large values of $N$, the ideal sensitivity is the Heisenberg limit divided by $\sqrt{2}$ for a wide range of $\mu$. For small values of $N$, the ideal sensitivity increases with $\mu$ until $\mu$ gets close to $\pi/2$. The signal and the noise level for $N = 5$, $\Gamma_0 = 0.1\chi$, and $\mu = 0.31\pi$ (arbitrarily chosen) are shown in Figure 2. The signal contrast is reduced as expected, while the noise is also suppressed at the operation points. The SE-induced reduction in the noise level increases with increasing $\mu$ and is maximized for $\mu = \pi/2$, corresponding to the SCSP. The signal and the noise level of the SCSP is shown in Figure 3. The SE-induced reduction in the noise level is an unexpectedly favorable discovery. Given the SCSP is identical to the GESP for $\mu = \pi/2$, the discussion regarding the SCSP will be implicitly included in discussions regarding the GESP, and the SCSP will not be mentioned separately, in the rest of this paper.

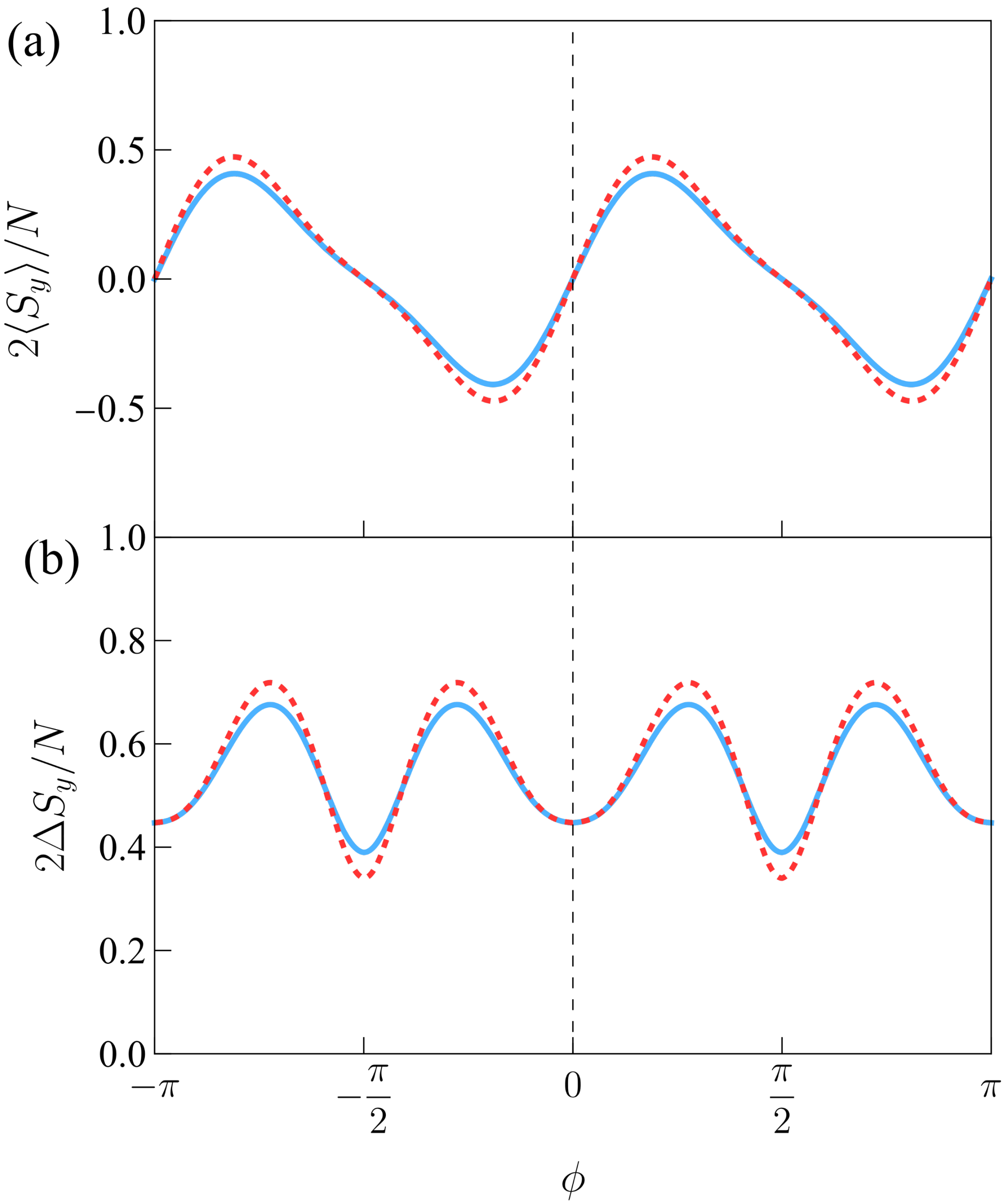

Figure 1. Signal and quantum noise level of the CESP as a function of the phase shift $\phi$ for $N=5$ and $\mu=\pi/6$ (the optimal value of $\mu$ for the CESP). The blue solid curves show the results for $\Gamma=0.2\chi$ and the red dashed for $\Gamma=0$. The vertical black dashed line marks the operation point. (a) Signal of the CESP, which is the expectation value of the measured operator $S_y$, normalized to the maximum amplitude $N/2$. (b) Quantum noise level of the CESP, which is the standard deviation of the measured operator $S_y$, normalized to the maximum noise level $N/2$.

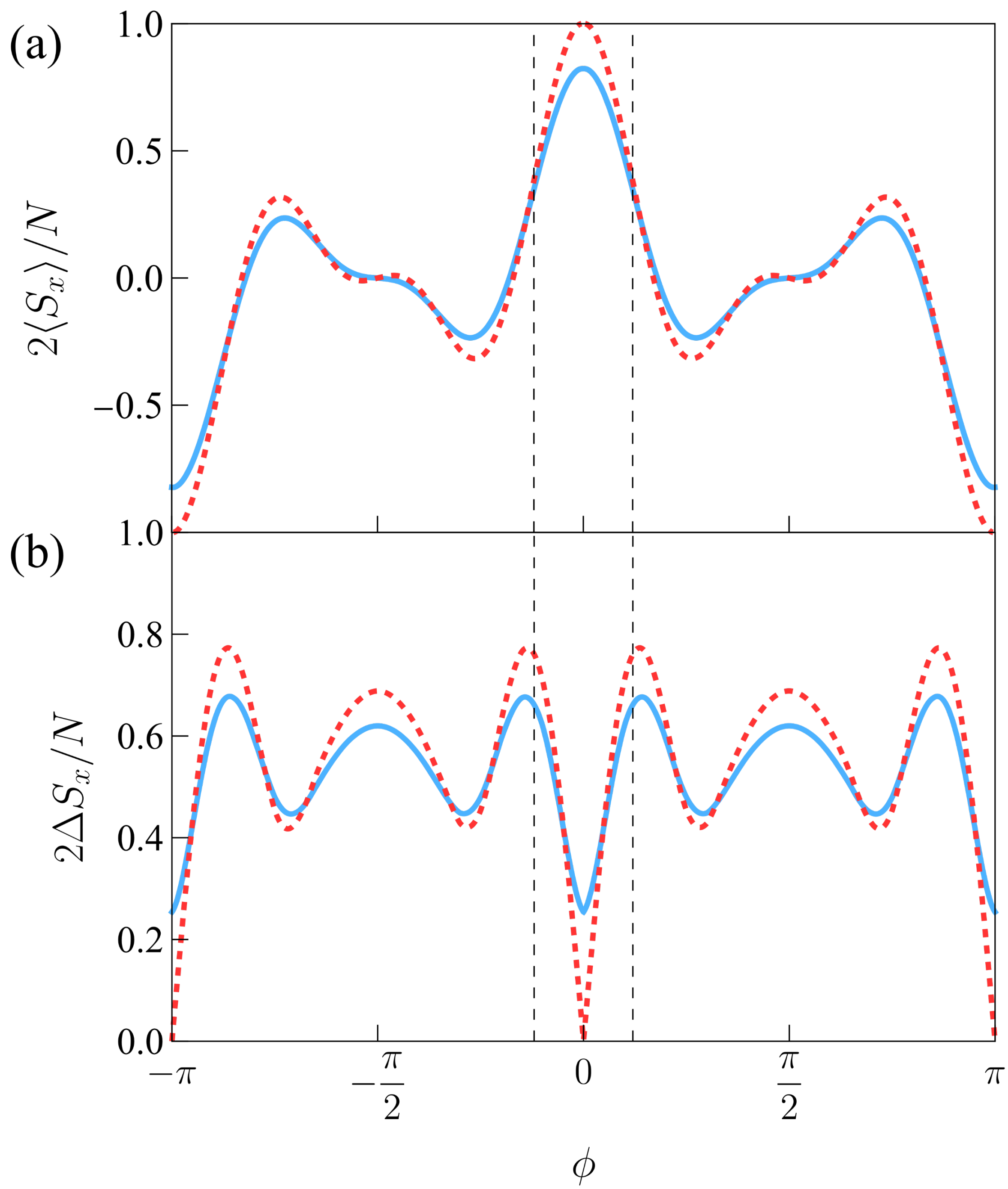

Figure 2. Signal and quantum noise level of the GESP as a function of the phase shift $\phi$ at $N=5$, $\Gamma_0=0.1\chi$, and $\mu=0.31\pi$ (a value arbitrarily chosen). The blue solid curves show the results for $\Gamma=0.2\chi$ and the red dashed for $\Gamma=0$. The vertical black dashed lines mark the operation points. (a) Signal of the GESP, which is the expectation value of the measured operator $S_x$, normalized to the maximum amplitude $N/2$. (b) Quantum noise level of the GESP, which is the standard deviation of the measured operator $S_x$, normalized to the maximum noise level $N/2$.

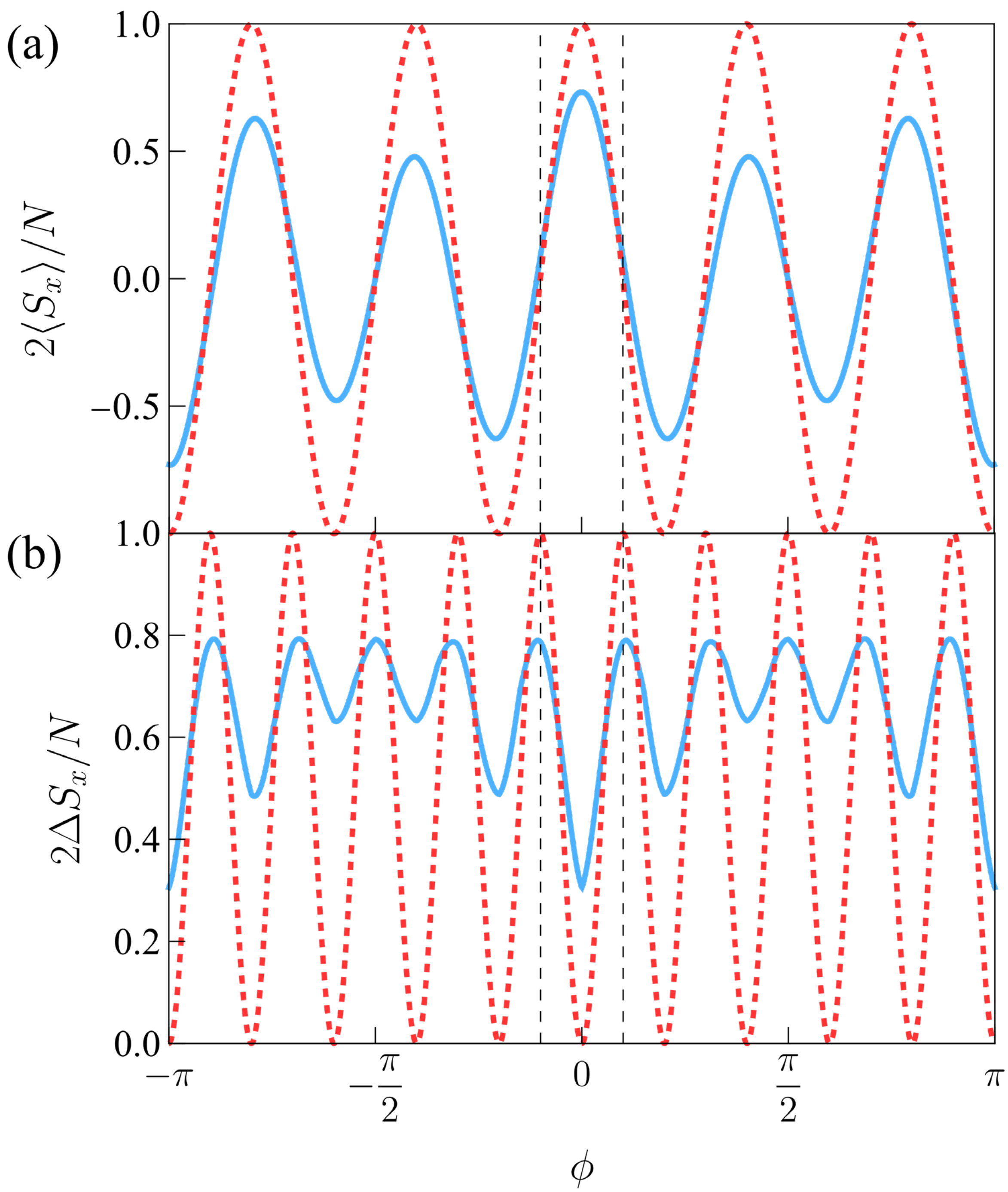

Figure 3. Signal and quantum noise level of the SCSP as a function of the phase shift $\phi$ for $N = 5$. The blue solid curves show the results for $\Gamma = 0.2\chi$ and the red dashed for $\Gamma = 0$. The vertical black dashed lines mark the operation points. (a) Signal of the SCSP, which is the expectation value of the measured operator $S_x$, normalized to the maximum amplitude $N/2$. (b) Quantum noise level of the GESP, which is the standard deviation of the measured operator $S_x$, normalized to the maximum noise level $N/2$.

To study the effect of SE on the signal contrast, it is necessary first to carefully define the signal contrast for the ESPs. For the GESP, the signal contrast is defined as $\left(\langle S_x\rangle_{\phi=0}-\langle S_x\rangle_{\phi=\pi}\right)/N$, which equals $2\langle S_x\rangle_{\phi=0}/N$ in the case considered in this paper. For the CESP, the signal contrast is defined as the ratio of the phase gradient $\partial\langle S_y\rangle/\partial\phi$ in the presence of SE to that in the ideal case at the operational point $\phi=0$. Recalling that for $\phi=0$, the ESPs reduce to Ramsey protocols without squeezing, the reduction in the signal contrast for the GESP is easy to derive. For unentangled atoms, the ensemble can be described by the average single-atom density operator, with the coherence (off-diagonal) density-matrix elements exponentially decaying at a rate $\Gamma/2$, reducing the signal-contrast to $e^{-\Gamma t/2}$ for Ramsey protocols without squeezing. Therefore, the signal contrast for the GESP $2\langle S_x\rangle_{\phi=0}/N$ is also reduced to $e^{-\Gamma t/2}$. Simulations with the values of $N$ permitted by available computational resources show that the signal contrast for the CESP is reduced to $e^{-3\Gamma t/4}$, regardless of the value of $N$. Although we have not found the analytical proof of this result, its validity can be expected for arbitrary $N$. Considering that the SE reduces the signal contrast less and suppresses the quantum noise for the GESP, it can be concluded that the GESP is more resistant to SE than the CESP. This is another advantage of the GESP that has not been realized formerly.

According to the expressions for $\chi$ and $\Gamma$ shown in the introduction, $\Gamma$ can be written as $\frac{\Gamma_0\chi}{(2g)^2}\frac{\left(4\delta^2+\kappa^2\right)}{\delta}$. The dephasing caused by cavity decay is another decoherence mechanism that depends on $\delta/\kappa$, which thus needs to be considered for optimizing the signal contrast. Accounting for both SE and cavity decay, the signal contrast is reduced to $\exp\left(-\frac{\kappa\mu}{\delta}-\frac{\Gamma_0\mu}{(2g)^2}\frac{\left(4\delta^2+\kappa^2\right)}{\delta}\right)$ [9], which takes its maximum value of $\exp\left(-4\mu\sqrt{C+1}/C\right)$ at $\delta=(\kappa/2)\sqrt{C+1}$, where $C\equiv(2g)^2/\kappa\Gamma_0$ is the single-atom cooperativity. Therefore, the single-atom cooperativity $C$ is the key factor determining the signal-contrast reduction induced by SE and cavity decay.

The value of $C$ for realistic parameters is calculated below. The single-photon Rabi frequency $2g$ can be expressed as the magnitude of the single-photon electric field $\sqrt{4\pi\hbar\omega/AL}$ (in Gaussian units) times the transition dipole moment $d$ divided by $\hbar$, which equals $d\sqrt{4\pi\omega/\hbar AL}$, where $\omega$ is the frequency of the cavity field, $A$ is the cross section at the waist of the cavity field, and $L$ is the length of the cavity field. The SE rate $\Gamma_0$ can be expressed as $4k^3d^2/3\hbar$, where $k=\omega/c$ is the wavenumber of the cavity field. The decay rate of the cavity induced by the transmission of the incident coupler is given by $\kappa = cT/L$, where $T$ is the transmittance of the incident coupler. Substituting these three expressions into the one for $C$ yields $C = 3\pi/ATk^2$. It should be noted that for a particular transition (fixed $k$), the single-atom cooperativity only depends on the transmittance of the incident coupler and the cross section at the waist of the cavity field. For the ${}^1S_0 \leftrightarrow {}^3P_1$ transition, $k = 2\pi \times 1.45\ \mu\text{m}^{-1}$. For $A = \pi(30\ \mu\text{m})^2$ and $T = 10^{-5}$, the single-atom cooperativity is about 4.0. For large $N$, the ideal sensitivity of the GESP remains nearly constant at the Heisenberg limit divided $\sqrt{2}$ in the interval $4\sqrt{2/N} \le \mu \le \pi/2 - \sqrt{2/N}$. In the absence of excess detection noise, it is obvious that the GESP should be implemented at $\mu = 4\sqrt{2/N}$ because the signal-contrast reduction caused by SE and cavity decay is minimized. However, in the presence of excess detection noise, the optimal $\mu$ may shift to a larger value within the interval. The case when the level of excess detection noise reaches $0.1N$ and $C = 4$ can be used as an example to estimate the optimal $\mu$. Given that the quantum noise is suppressed by SE for the GESP, the SE-induced reduction in the signal contrast alone provides an upper limit of the sensitivity degradation caused by SE. The lower limit of the GESP sensitivity normalized to the Heisenberg limit is given by

$$(N\Delta\phi)^{-1} = \frac{|\partial S/\partial\phi|}{\sqrt{\Delta S_Q^2 + \Delta S_E^2}} = \frac{\exp\left(-4\mu\sqrt{C+1}/C\right)N\sin\mu\Big/\sqrt{2}}{\sqrt{N^2\sin^2\mu + 0.01N^2}} = \frac{e^{-\sqrt{5}\mu}\sin\mu}{\sqrt{2\sin^2\mu + 0.02}} \tag{8}$$

This normalized sensitivity takes its maximum value of 0.47 at $\mu = 0.04\pi$. Figure 4 shows the sensitivity as a function of $\mu$ for $C = 4$ (green) and $C = 1$ (orange) in the presence of excess detection noise at a level of $0.1N$.

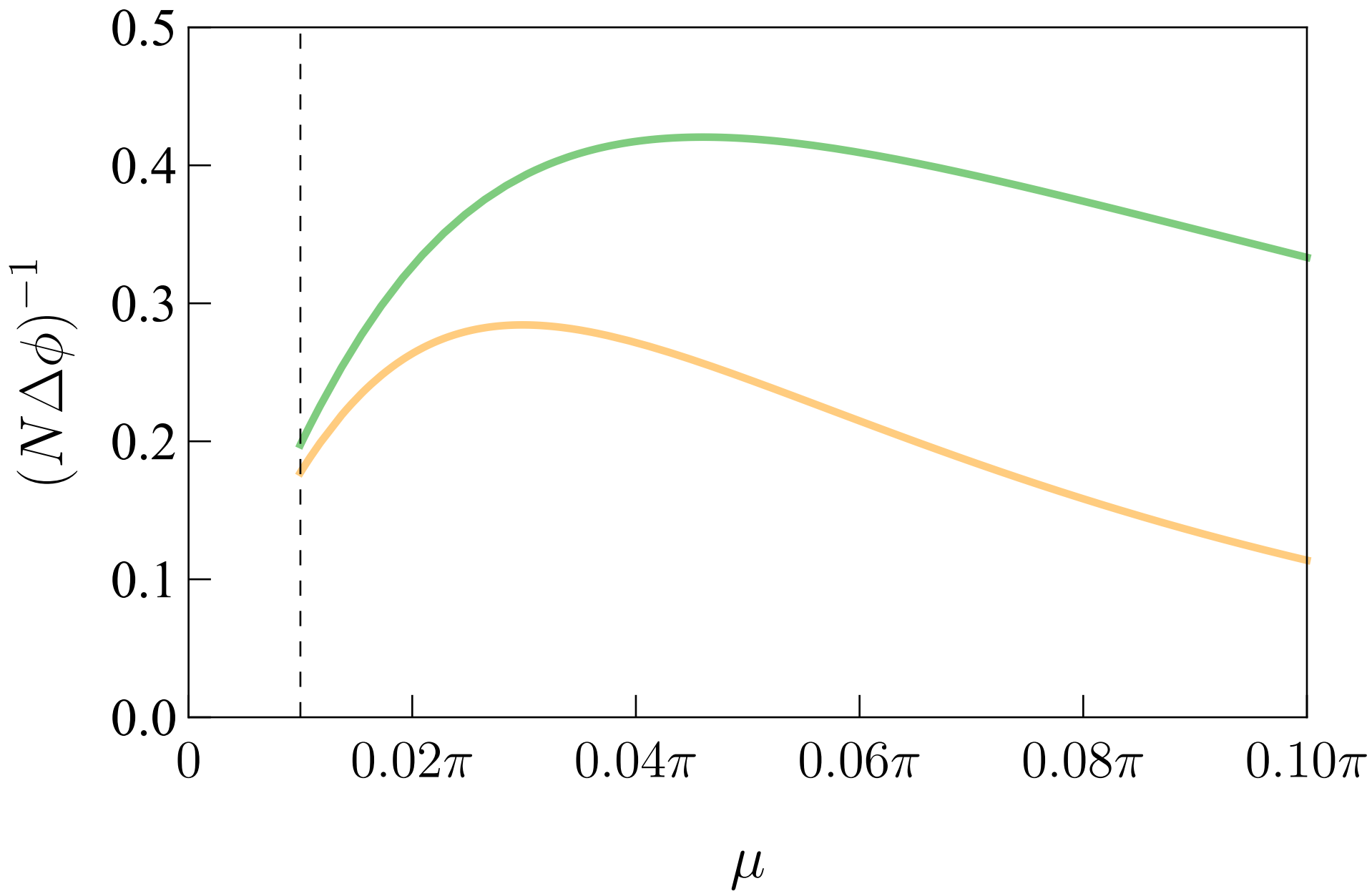

Figure 4. Lower limit of the GESP sensitivity normalized to the Heisenberg limit as a function of $\mu$ for $C = 4$ (green) and $C = 1$ (orange) in the presence of excess detection noise at a level of $0.1N$ .

## 4. Conclusion

OATS is a technique widely used for enhancing the sensitivity of quantum sensors. For atomic sensors, SE in the OATS operations is an important imperfection, which, however, is prohibitively difficult to study. SE can transfer atoms to all the Zeeman substates, making the number of relevant collective states excessively large. Earlier studies only focused on the SE-induced increase in the quantum noise along the short axis of the noise ellipse in the moderate squeezing regime because the suppression of the quantum noise was the original purpose of spin squeezing. These studies are limited in three aspects. First, the complexity introduced by $m_F \neq 0$ Zeeman substates is completely ignored. Second, these results are not applicable to ESPs because the ESPs are all implemented in the oversqueezed regime. Third, the SE-induced reduction in the signal contrast is not included. In this paper, we address the three limitations as follows. To avoid the complexity introduced by $m_F \neq 0$ Zeeman substates, we focus on a relatively simple isotope, namely $^{88}$Sr, which only has one Zeeman substate in each of the two ground states. Nevertheless, studying the effect of SE is still challenging because SE populates all collective states, either symmetric or asymmetric, thus putting the ensemble into a mixed state. To study the effect of SE in the whole interval $0 \le \mu \le \pi/2$ rather than only small values of $\mu$, we establish a general model without making assumptions on the value of $\mu$ and conduct numerical simulations for values of $N$ permitted by available computational resources as an aid when analytical derivation is intractable. In addition to the effect of SE on the quantum noise, we also investigate the SE-induced reduction in the signal contrast in detail. The conclusions of our study are summarized below. For the GESP employing $^{88}$Sr, the signal contrast is reduced to $e^{-\Gamma t/2}$, which is the same as that for a Ramsey protocol without squeezing. For the CESP employing $^{88}$Sr, the signal contrast is reduced to $e^{-3\Gamma t/4}$. The quantum noise at the operation points is suppressed by SE for the GESP and remains unchanged for the CESP. The SE-induced suppression of quantum noise constitutes a previously unrecognized effect that challenges both earlier conclusions and intuitive expectations. Given that for the GESP, SE reduces the signal contrast less and suppresses the quantum noise, we conclude that the GESP is more resistant to SE than the CESP. This is another advantage of the GESP that has not been realized formerly. Therefore, the reduction in signal contrast alone provides an upper limit of the SE-induced sensitivity degradation. Accounting for both SE and cavity decay, the signal contrast takes its maximum value of $\exp\left(-4\mu\sqrt{C+1}/C\right)$ at $\delta = (\kappa/2)\sqrt{C+1}$. When the level of excess detection noise reaches $0.1N$, the lower limit of the GESP sensitivity normalized to the Heisenberg limit takes its maximum value of 0.47 at $\mu = 0.04\pi$ for $C = 4$.

**Acknowledgement:**

This work has been supported equally in parts by the Department of War Center of Excellence in Advanced Quantum Sensing under Army Research Office grant number W911NF202076, and NASA NIAC grant number 80NSSC25K8008.